\documentclass[sigconf]{acmart}
\usepackage{amsmath}
\usepackage{graphicx}
\usepackage{hyperref}
\hypersetup{
    colorlinks=true,
    linkcolor=blue,
    filecolor=magenta,      
    urlcolor=cyan,
    }

\usepackage{float}
\usepackage{url}

\usepackage{booktabs}
\usepackage{mdframed} 
\usepackage{relsize}

\usepackage{enumitem}

\AtBeginDocument{%
  }

\setcopyright{acmlicensed}
\copyrightyear{2025}
\acmYear{2024}
\acmDOI{XXXXXXX.XXXXXXX}

\acmISBN{978-1-4503-XXXX-X/18/06}


\begin{document}

\title[LLM-Enhanced Framework for Improving Personalized Recommendations]{A Language-Driven Framework for Improving Personalized Recommendations: Merging LLMs with Traditional Algorithms}

\author{Aaron Goldstein}
\email{n01421643@unf.edu}
\affiliation{%
  \institution{University of North Florida}
  \city{Jacksonville}
  \state{Florida}
  \country{USA}
}

\author{Ayan Dutta}
\email{a.dutta@unf.edu}
\affiliation{%
  \institution{University of North Florida}
  \city{Jacksonville}
  \state{Florida}
  \country{USA}
}

\renewcommand{\shortauthors}{Goldstein and Dutta}


\begin{abstract}
Traditional recommendation algorithms are not designed to provide personalized recommendations based on user preferences provided through text, e.g., "I enjoy light-hearted comedies with a lot of humor". Large Language Models (LLMs) have emerged as one of the most promising tools for natural language processing in recent years. This research proposes a novel framework that mimics how a close friend would recommend items based on their knowledge of an individual's tastes. We leverage LLMs to enhance movie recommendation systems by refining traditional algorithm outputs and integrating them with language-based user preference inputs. We employ Singular Value Decomposition (SVD) or SVD++ algorithms to generate initial movie recommendations, implemented using the Surprise Python library and trained on the MovieLens-Latest-Small dataset. We compare the performance of the base algorithms with our LLM-enhanced versions using leave-one-out validation hit rates and cumulative hit rates. Additionally, to compare the performance of our framework against the current state-of-the-art recommendation systems, we use rating and ranking metrics with item-based stratified 0.75 train, 0.25 test split. 
Our framework can generate preference profiles automatically based on users' favorite movies or allow manual preference specification for more personalized results. Using an automated approach, our framework overwhelmingly surpassed SVD and SVD++ on every evaluation metric used (e.g., improvements of up to $\sim 6$x in cumulative hit rate, $\sim 3.7$x in NDCG, etc.), albeit at the cost of a slight increase in computational overhead. 
\end{abstract}


\maketitle


\section{Introduction}
Recommendation systems~\cite{resnick1997recommender} are integral parts of modern-day online experiences ranging from e-commerce to OTT platforms~\cite{aggarwal2016introduction,chiang2024recommender,roy2022systematic}. Given the recent rise in capabilities of Large Language Models (LLMs)~\cite{achiam2023gpt,katz2024gpt,nori2023capabilities}, we can utilize them to understand natural language-based user preferences before providing final recommendations to customers. Certain companies such as Amazon are already giving such options on their Fire devices showing suggestions like "Show me Al Pacino's action movies". Despite the already exciting number of studies that focus on combining LLMs and recommendation systems (RS), extensively reviewed by \cite{wu2024survey}, there is still much room for further advancement and innovation.

As outlined by Wu et al.~\cite{wu2024survey}, LLM-based recommendation approaches can be categorized into three main paradigms: (1) LLM Embeddings + RS, where LLMs serve as feature extractors, converting features into embeddings which are then used by RS systems to improve their performance~\cite{wu2021empowering}; (2) LLM Tokens + RS, where LLMs generate tokens that capture nuanced user preferences and item characteristics, this tokenized form can be used for recommendation refinement tasks as in the framework we present here~\cite{liu2024once}; and (3) LLM as RS, where LLMs directly generate recommendations~\cite{dai2023uncovering}. Our framework primarily follows the second paradigm, using (non-tuning) LLM-generated semantic insights to improve and refine the outputs of any traditional algorithms which provides top $N$ output recommendations.

To incorporate natural language-based user preferences into the final recommendations, in this paper, we propose using an LLM-based architecture. The LLM is used in four key places: 1) in movie information retrieval through fuzzy matching when IMDb IDs are outdated or broken, 2) if movie descriptions are not available online after retrieval attempts, we use the LLM to generate concise descriptions, 3) in automatic generation of user preference profiles when users cannot or choose not to provide preferences manually, and 4) in similarity scoring of titles recommended by existing state-of-the-art SVD-based recommendations~\cite{koren2009matrix}. The similarity scoring system helped us to fine-tune the SVD or SVD++~\cite{koren2008factorization} recommendations and provide users with better-quality recommendations. The scoring system is designed by us and provided to the LLM as an example to begin with. The LLM learned from those examples and provided recommendations based on the scoring system (see Section \ref{sec:similar_scoring}). 

We have implemented the proposed framework on the MovieLens-Latest dataset and compared hit rate as well as various ranking metrics such as Normalized Discounted Cumulative Gain (NDCG) and Mean Average Precision (MAP), among others. Results show that our presented LLM-enhanced recommendations comprehensively outperform the base SVM and SVD++ recommendations in all the used metrics. 

The main contributions of our paper can be summarized as follows.
\begin{itemize}
    \item Our novel framework graciously handles user preferences provided to it via natural language. Furthermore, we account for popularity preferences in prompting for bias mitigation.
    
    \item We present a novel framework that leverages LLM to fine-tune recommendations provided by a base algorithm to the users. LLM is also used in a novel fashion to create descriptions of movies that are unavailable or in information retrieval when IMDb IDs are broken.

    \item Results show that the proposed framework thoroughly outperforms the base algorithms in terms of hit rate and four other ranking metrics, including NDCG and MAP.
\end{itemize}


\section{Methodology}
\subsection{Preliminaries}
\begin{figure*}
    \centering
    \includegraphics[width=0.75\textwidth]{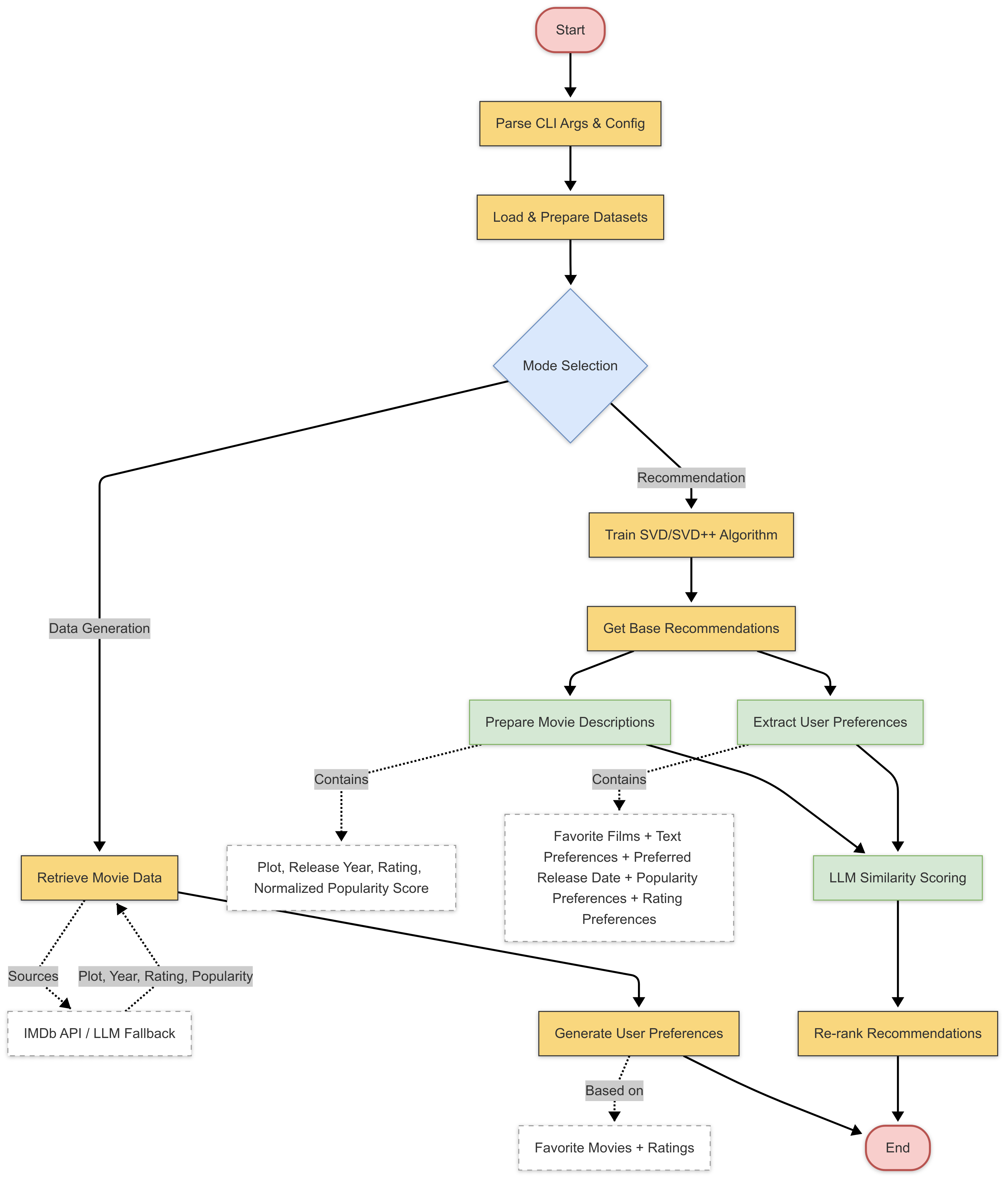}
    \caption{Framework Architecture Flowchart showing data generation and recommendation paths}
    \label{fig:system-flowchart}
\end{figure*}

This research proposes a framework that leverages the capabilities of Large Language Models to enhance recommendation systems by refining the recommendations of traditional recommendation algorithms such as SVD and SVD++. This approach builds on the foundational work of the studies discussed in the related works section. The proposed framework overview is illustrated in Fig. \ref{fig:system-flowchart}.

\noindent
\textbf{Data Preparation.}
In our hit rate and ranking metric experiments where we generate user preference profiles for all dataset users, we utilize the MovieLens Latest Small (100k dataset)~\cite{harper2015movielens}, which provides a comprehensive set of user ratings for recommendations. The stable 100k version from 1998 is outdated and contains old and no longer functioning IMDb links. This is why we chose the Latest Small Version (last updated 9/2018) that is more up to date with current information (i.e, more likely to contain accurate IMDb links), and contains newer and relevant movies. 

This dataset's CSV files are loaded into Pandas DataFrames using Python, and we create mapping dictionaries between MovieLens Movie IDs and IMDb IDs using the \path{links.csv} included in the MovieLens dataset. This facilitates seamless integration with IMDb data, enabling the retrieval of additional movie information, including plots and reviews if needed. For our study, using Latest-Small instead of MovieLens-32M was necessary for fully comparing hit rate between LLM Enhanced and the corresponding base traditional algorithms, as generating similarity scores and user preference profiles for all users in the full MovieLens dataset would be unfeasible due to our local resource capacity. 

\smallskip
\noindent
\textbf{Traditional Recommendation Algorithms.} We employ both SVD and SVD++ algorithms to generate initial movie recommendations. SVD and SVD++ are matrix factorization techniques that uncover latent factors in user-item interactions, making them effective for handling sparse data~\cite{koren2009matrix}. For calculating hit rate, the algorithms are trained on the MovieLens-Latest-Small dataset, and we compare their performance with our LLM-enhanced versions using hit rate~\cite{AcharyaLLM2023} and other ranking metrics such as NDCG~\cite{zangerle2022evaluating}. 

\smallskip
\noindent
\textbf{LLM-enhanced Framework Description Generation.}
The integration of generating descriptions is inspired by Acharya et al. \cite{AcharyaLLM2023}. We use LLMs to generate movie descriptions through few-shot prompting when official descriptions can not be retrieved from IMDb. This approach reduces reliance on web scraping and mitigates personal biases in descriptions, aligning with Acharya et al.'s \cite{AcharyaLLM2023} findings that LLM-generated descriptions can achieve performance close to manually curated ones. 

Our system locally uses the Phi-4-Q6\_K quantization model. This decision allows the model to run efficiently on an RTX 4090 Laptop GPU with 16 GB VRAM, providing a good balance between model capability and resource constraints while improving recommendation quality. Phi-4 is known to be particularly good at math related reasoning, surpassing even much larger models like LLama-3.3 70B Instruct, Claude 3.5 Sonnet, GPT 4o, and GPT 4o-mini on math related problems which makes it particularly suitable for our use case heavily dependent on numerical scores~\cite{abdin2024phi}.

We also implement LLM fuzzy matching to search the IMDb Cinemagoer API for movie data when presented with outdated or broken IMDb IDs in the dataset, addressing a common issue in the MovieLens dataset where some IDs are no longer valid. This reduces the amount of cases where descriptions can not be found. System users can also match movies that they want to rate by searching through the IMDb database, then convert matches to MovieLens movie IDs.

An example of the description generation prompt can be seen in Figure \ref{fig:Description Generation Prompt}. Few-shot prompting allows the model to learn from a limited number of examples, making it adaptable to new tasks with minimal data. All few shot example movie descriptions in the example prompts shown are retrieved from IMDb plot summaries or trailer descriptions. Our constructed prompts for the LLM using few-shot examples, guides the model in generating logically sound if not accurate descriptions depending on the training dataset.

\begin{figure*} \begin{mdframed}[linewidth=1pt] \begin{quote}\footnotesize 

\textbf{System Prompt: }

You are a helpful assistant that generates concise movie descriptions. Do not use newlines in your response. The examples provided are for context only and should not appear in your output. Return only the description.

\textbf{User Prompt:}

Example 1:
Movie title: Inception
Description: A thief who steals corporate secrets through dream-sharing technology is tasked with implanting an idea into a CEO's mind.

Example 2:
Movie title: The Matrix
Description: A computer hacker discovers his reality is an illusion and joins rebels to fight its controllers.

Generate a description for this movie:

Movie title: Interstellar

Description: \textcolor{blue}{[The LLM will generate the description for Interstellar here - following the pattern established by the few-shot examples above.]}
\end{quote} 
\end{mdframed} 
\caption{A sample description generation prompt with few-shot example descriptions retrieved from IMDb} 
\label{fig:Description Generation Prompt}
\end{figure*}

\subsection{Similarity Scoring and Personalized Recommendations}
\label{sec:similar_scoring}
Inspired by Maragheh et al. \cite{yousefi2024llm}, our system uses LLMs to generate similarity scores between user preferences and movie titles, followed by their descriptions. It leverages an LLM to evaluate the similarity between a user's preferences and a set of movie descriptions of length \( N \times T \times 10^M \), sorted by the highest rating to the lowest that the base algorithm (e.g., SVD, SVD++) predicts a user will rate a movie based on their past ratings. \( N \) is the final number of movies to recommend to the user while \( M \) can be modified for more accurate results, at the cost of performance. \(T\) is a tuning factor in the range $(0, 1]$ for more fine grained control. If we wanted a search count (the size of the candidate pool of top base recommendations to re-rank) of $100$ for an $N = 5$, the search count equation would intuitively be \(5 \times 0.2 \times 10^2\), where \(T\) = $0.2$ and \(M\) = $2$.

In our testing, we used a search count of $100$ for each $N$ ($N$@1, $N$@5, and $N$@10), meaning we select the top 100 base recommendations for each user, assign a similarity score to each, and then re-rank them according to those scores. The top \( N \) movies with the highest similarity scores are returned as personalized recommendations. This approach builds on Maragheh et al.'s \cite{yousefi2024llm} use of LLMs to generate aspect embeddings that refine recommendation systems by providing contextual information and aligning recommendations with user intention.

\smallskip
\noindent
\textbf{Similarity Scoring Considerations. }
Our similarity scoring process incorporates several important factors beyond just text preferences and movie descriptions:

\begin{itemize}
    \item \textit{IMDb Ratings:} When available, each movie's IMDb rating (on a scale of 0-10) is included in the prompt.
    \item \textit{Popularity Scores:} A normalized popularity score (0-100) derived from IMDb vote counts is included when available. This score is calculated using linear interpolation where 1 million votes corresponds to a score of 100 as follows. 
    \begin{equation}
        \text{normalized\_score} = \min(100, \max(0, 100 \times \frac{\text{votes}}{1{,}000{,}000}))
        \label{eq:score}
    \end{equation}
    \item \textit{Rating Preferences:} For users who tend to favor highly-rated content, we conditionally include the statement ``I prefer movies with high IMDb ratings" in their preference profile.
    \item \textit{Popularity Preferences:} Similarly, for users who gravitate toward mainstream titles, we include ``I prefer popular/trending movies" in their profile.
    \item \textit{Release Date Range:} A preferred release date range is determined based on the user's rated movies, rounding down to the first year of the decade for the minimum year and up to the first year of the next decade (or current year, whichever is smaller) for the maximum year.
\end{itemize}

Both rating and popularity preferences are determined automatically during the generation of user preference profiles. If the average IMDb rating of a user's favorite movies is $\geq 7.0/10$, the rating preference is enabled. Similarly, if the average normalized popularity score is $\geq 80/100$ (Eq. \ref{eq:score}), the popularity preference is enabled. These elements appear conditionally in the LLM prompt only when relevant to that particular user's profile. These factors were implemented to remove obscure titles being recommended to users who might prefer more mainstream titles.

Our framework can generate these preference profiles for all users in the dataset automatically based on their favorite movies, which is how we scale to use the framework on an entire dataset. Alternatively, users can specify all their preferences manually if desired. Due to performance considerations, the framework should be applied to users on an as-needed basis, allowing streaming service platforms to personalize recommendations for users who wish to provide their preferences.

\smallskip
\noindent
\textbf{Similarity Scoring Prompt.}
To compute the similarity between a user's preferences and a movie's description, we construct a prompt that provides the user's input and the movie's details to the LLM and ask the LLM to return a similarity score between $-1.0$ and $1.0$, where $-1.0$ means the movie goes completely against the user's preferences, $0$ means neutral or not enough information, and $1.0$ is a perfect match. We use regex expression searching to extract these similarity scores from LLM responses, ensuring consistent output format and stability. The prompt includes few-shot examples to guide the LLM's response, and is specifically formatted to accommodate the Phi-4 model's input structure with properly formatted system and user messages. The combination between few shot examples and regex expression searching allows even small models to quickly learn the task and provide accurate responses. An example of the prompt is seen as follows in Fig. \ref{fig:Sim_Scoring_Prompt} without the special Phi-4 model formatting for generalization. 

\begin{figure*} 
\begin{mdframed}[linewidth=1pt] \begin{quote}\footnotesize 

\textbf{System:}

You are a movie recommendation assistant. Your task is to evaluate how well a movie description aligns with a user's stated preferences and their favorite movies. Always respond with a number between -1.0 and 1.0, where:
-1.0 means the movie goes completely against their preferences,
0 means neutral or there isn't enough information,
1.0 is a perfect match. You must respond with only the number, without any additional text or formatting under all circumstances.

\textbf{User:}

Example 1:

User input: I love science fiction with deep philosophical themes. 

I prefer movies with high IMDb ratings.

User's favorite movies:

Movie title: The Matrix (1999)

Movie title: Blade Runner (1982)

Movie title: Interstellar (2014)

Preferred Release Date Range: I prefer movies released between 1980 and 2020.

New movie to evaluate:

Movie title: Inception (2010)

IMDb Rating: 8.8/10

Popularity Score: 93/100

Movie description: A thief who steals corporate secrets through the use of dream-sharing technology is given the inverse task of planting an idea into the mind of a C.E.O., but his tragic past may doom the project and his team to disaster.

Rate how likely you think the movie aligns with the user's interests (respond with a number in range [-1, 1]):

0.9

Example 2:

User input: I enjoy light-hearted comedies with a lot of humor. 

I prefer popular/trending movies.

User's favorite movies:

Movie title: The Hangover (2009)

Movie title: Superbad (2007)

Movie title: Step Brothers (2008)

Preferred Release Date Range: I prefer movies released between 2000 and 2010.

New movie to evaluate:

Movie title: The Dark Knight (2008)

IMDb Rating: 9.0/10

Popularity Score: 98/100

Movie description: Set within a year after the events of Batman Begins (2005), Batman, Lieutenant James Gordon, and new District Attorney Harvey Dent successfully begin to round up the criminals that plague Gotham City, until a mysterious and sadistic criminal mastermind known only as "The Joker" appears in Gotham, creating a new wave of chaos.

Rate how likely you think the movie aligns with the user's interests (respond with a number in range [-1, 1]):

-0.7

Example 3:

User input: I am fascinated by historical documentaries.

User's favorite movies:

Movie title: They shall not grow old (2018)

Movie title: Apollo 11 (2019)

Movie title: 13th (2016)

Preferred Release Date Range: I prefer movies released between 2010 and 2020.

New movie to evaluate:

Movie title: The Lord of the Rings: The Fellowship of the Ring (2001)

IMDb Rating: 8.8/10

Popularity Score: 95/100

Movie description: A meek Hobbit from the Shire and eight companions set out on a journey to destroy the powerful One Ring and save Middle-earth from the Dark Lord Sauron.

Rate how likely you think the movie aligns with the user's interests (respond with a number in range [-1, 1]):

-0.5

Now, respond to the following prompt:

User input: I am drawn to science fiction and fantasy stories that are well written.

I prefer movies with high IMDb ratings.

I prefer popular/trending movies.

User's favorite movies:

Movie title: The Lord of the Rings: The Two Towers (2002)

Movie title: Rogue One: A Star Wars Story (2016)

Movie title: WALL·E (2008)

Preferred Release Date Range: I prefer movies released between 2000 and 2020.

New movie to evaluate:

Movie title: Interstellar (2014)

IMDb Rating: 8.6/10

Popularity Score: 91/100

Movie description: When Earth becomes uninhabitable in the future, a farmer and ex-NASA pilot, Joseph Cooper, is tasked to pilot a spacecraft, along with a team of researchers, to find a new planet for humans.

Rate how likely you think the movie aligns with the user's interests (respond with a number in range [-1, 1]):

\end{quote} 
\end{mdframed} 
\caption{Similarity Scoring Prompt with preference statements, preferred release range, favorite movies, IMDb ratings and popularity scores} 
\label{fig:Sim_Scoring_Prompt}
\end{figure*}


\section{Experiments and Results}
The code is anonymously available at \cite{4openAnonymousGithub}.

\subsection{Implementation Details}
We designed the proposed system to work in three different modes: development, production, and data generation. Development mode works with a locally running the LLM model that exposes an OpenAPI-compatible API. In this mode, we tested with a small but powerful \verb|Phi-4-Q6_K| quantization model. The Q6\_K quantization allows us to easily offload all model layers to an RTX 4090 Laptop GPU with 16 GB VRAM while still having 4096 tokens of context, which is more than enough for our purposes. The production mode allows the system to use larger LLM models by outsourcing the LLM computation to OpenAI, by default, utilizing their current fast, and cheap GPT-4o-mini model if so desired, though we recommend local execution due to latency and the sheer number of API calls made by the system. Our framework generates many prompts with identical instruction segments and few-shot examples, for this reason, for local execution we elected to use KoboldCPP, a sophisticated application for running local GGUF models, thanks to its Context Shifting implementation which significantly reduces token usage and computational overhead while maintaining the quality of similarity assessments. Utilizing KV cache shifting, KoboldCPP's Context Shifting feature automatically removes old tokens from context and adds new ones without any reprocessing, allowing us to use long instructional sequences and many few shot examples while still only processing approximately $100$ tokens for the vast majority of similarity score calculations instead of approximately $1000$-$1500$ tokens each time, helping mitigate the potential performance drawback from such techniques.

We designed generate data mode to be a comprehensive data retrieval solution for the framework, gathering data such as movie descriptions from IMDb through the Cinemagoer library, utilizing MovieLens-Latest-Small's MovieLens IDs to IMDb IDs mapping in the \path|links.csv| file to accurately locate movies. 
We also implemented fuzzy matching to search the IMDb Cinemagoer API when we need to locate movies with outdated or broken IMDB IDs in the \path|links.csv| file. When a new user is providing their favorite movies to the framework to receive recommendations, they can also manually search through the IMDb catalog to identify their favorite movies, where we then convert the IMDb IDs back to MovieLens IDs. If the system cannot find a description for a movie in the MovieLens dataset, for example, because of an outdated MoveLens ID to IMDb ID link, the system will generate a description for the movie using an LLM. 

We currently handle description generation locally due to the sheer scale of the dataset and to reduce the number of calls to OpenAI for cost savings. Our system caches retrieved and generated movie descriptions to optimize performance. Our caching strategy reduces the need for repeated API calls to IMDb, and the OpenAI-compatible API, improving system efficiency.

By building on the methodologies of \cite{AcharyaLLM2023,yousefi2024llm}, our proposed framework of LLM-enhanced algorithm for recommendations demonstrates the potential of LLMs to enhance and build upon traditional recommendation system algorithms such as SVD and SVD++ (implemented using Surprise Python library~\cite{hug2020surprise}), providing personalized and context-aware movie suggestions that align with user preferences. Our work continues the trend of integrating LLMs at various stages of the recommendation process, from description generation to similarity scoring, ultimately aiming to tailor recommendations based on user preferences and intentions -- this helps us to recommend items users actually enjoyed rather than items they have simply interacted with.
 
\subsection{Comparison using Hit Rate}
Using our framework, we generated preference profiles automatically for all users in the MovieLens-Latest-Small dataset based on their favorite movies, incorporating normalized popularity scores, rating preferences, and release date preferences. We then calculated hit rates for both traditional algorithms (SVD and SVD++) as well as for their LLM-enhanced versions using our proposed technique. We tested with N@1, N@5, and N@10 for both hit rate and cumulative hit rate, recognizing that up to first ten recommendations are the most crucial for user satisfaction. For normal Hit Rate, we calculate hits by dividing the number of times we predict the left out movie across all users by the length of the test set (i.e., number of left out movies). In Cumulative Hit Rate, we filter the test dataset to only have left out movies with an original rating $\geq 4.0$ (out of $5$ as MovieLens rating are scaled between $[0.5, 5.0]$), then calculate Cumulative Hit Rate as the number of times we predict the left out movie in the filtered dataset divided by the new length of the filtered dataset.

It is key to mention that we ensure we do not include the left out test set movie as a favorite movie for each user in their user preference profile that is provided to the LLM, as we felt this might be unfair when asking the LLM to rate the similarity of that movie to the user. Our results, shown in Tables \ref{tab:svd_comparison}, \ref{tab:svdpp_comparison}, and \ref{tab:averages} demonstrate significant improvements with the LLM-enhanced approach.

\begin{table*}
\centering
\caption{SVD LLM vs. SVD Comparison using Phi-4 on ML-Latest-Small}
\label{tab:svd_comparison}
\begin{tabular}{lccc}
\toprule
\textbf{Metric} & \textbf{SVD LLM} & \textbf{SVD} & \textbf{Improvement Ratio} \\
\midrule
Hit Rate N@1 & \textbf{0.008197} & 0.001639 & 5.001 \\
Hit Rate N@5 & \textbf{0.015847} & 0.006557 & 2.417 \\
Hit Rate N@10 & \textbf{0.018579} & 0.009290 & 1.999 \\
Cumulative Hit ($\geq$ 4.0) N@1 & \textbf{0.011662} & 0.001944 & 5.998 \\
Cumulative Hit ($\geq$ 4.0) N@5 & \textbf{0.024295} & 0.008746 & 2.778 \\
Cumulative Hit ($\geq$ 4.0) N@10 & \textbf{0.029155} & 0.011662 & 2.500 \\
\bottomrule
\end{tabular}
\end{table*}

\begin{table*}
\centering
\caption{SVD++ LLM vs. SVD++ Comparison using Phi-4 on ML-Latest-Small}
\label{tab:svdpp_comparison}
\begin{tabular}{lccc}
\toprule
\textbf{Metric} & \textbf{SVD++ LLM} & \textbf{SVD++} & \textbf{Improvement Ratio}  \\
\midrule
Hit Rate N@1 & \textbf{0.004372} & 0.001639 & 2.668 \\
Hit Rate N@5 & \textbf{0.010929} & 0.003279 & 3.333 \\
Hit Rate N@10 & \textbf{0.014208} & 0.008197 & 1.733 \\
Cumulative Hit ($\geq$ 4.0) N@1 & \textbf{0.007890} & 0.002959 & 2.667 \\
Cumulative Hit ($\geq$ 4.0) N@5 & \textbf{0.018738} & 0.004931 & 3.800 \\
Cumulative Hit ($\geq$ 4.0) N@10 & \textbf{0.024655} & 0.011834 & 2.084 \\
\bottomrule
\end{tabular}
\end{table*}

\begin{table*}[t]
\centering
\caption{Performance Averages Across Different N Values using Phi-4 on ML-Latest-Small}
\label{tab:averages}
\begin{tabular}{lcccc}
\toprule
\textbf{Metric} & \textbf{SVD} & \textbf{SVD LLM} & \textbf{SVD++} & \textbf{SVD++ LLM} \\
\midrule
Average Hit Rate & 0.005829 & \textbf{0.014208} & 0.004372 & 0.009836 \\
Average Cumulative Hit Rate & 0.007451 & \textbf{0.021704} & 0.006575 & 0.017094 \\
\bottomrule
\end{tabular}
\end{table*}

The results show that our LLM-enhanced approach consistently outperforms the base algorithms across all hit rate metrics. For SVD, we observe significant improvements across all N values, with the most dramatic increase at N@1 where the LLM-enhanced version achieves a $5$x improvement in hit rate ($0.008197$ vs $0.001639$) and a nearly $6$x improvement in cumulative hit rate ($0.011662$ vs $0.001944$). Similarly, for SVD++, the LLM-enhanced version shows substantial gains, particularly for N@5 cumulative hit rate, where we observe a $3.8$x improvement ($0.018738$ vs $0.004931$). On average, as shown in Table \ref{tab:averages}, the SVD LLM achieves a $2.44$x improvement in hit rate and a $2.91$x improvement in cumulative hit rate over standard SVD. These results demonstrate that the proposed LLM-enhanced framework provides a better understanding of individual users' preferences, resulting in more personalized and relevant recommendations.


\begin{table*}[t]
\centering
\caption{Average Recommendation Time Per User Comparison (sec.) with a Sample Size of 100 Users using Phi-4}
\label{tab:average_timing_results}
\begin{tabular}{lcc}
\toprule
\textbf{Metric} & \textbf{MovieLens Latest Small} & \textbf{MovieLens 32M} \\
\midrule
SVD Time & \textbf{0.037} & \textbf{1.756} \\
SVD LLM Time & 23.190 & 23.342 \\
Total Time & 23.227 & 25.097 \\
\bottomrule
\end{tabular}
\end{table*}

\subsection{Ranking Metrics Evaluation}

\begin{table*}[t]
\centering
\caption{Ranking Metrics Comparison using Phi-4 on ML-Latest-Small Dataset (N=10)}
\label{tab:ranking_metrics}
\begin{tabular}{lcccc}
\toprule
\textbf{Algorithm} & \textbf{NDCG@10} & \textbf{MAP@10} & \textbf{Precision@10} & \textbf{Recall@10} \\
\midrule
SVD & 0.090 & 0.042 & 0.078 & 0.026 \\
SVD LLM & 0.157 & 0.080 & \textbf{0.130} & \textbf{0.051} \\
SVD++ & 0.090 & 0.045 & 0.081 & 0.023 \\
SVD++ LLM & \textbf{0.160} & \textbf{0.085} & \textbf{0.130} & 0.050 \\
FastAI~\cite{FastaiMakingNeural} & 0.135 & 0.023 & 0.114 & 0.047 \\
\bottomrule
\end{tabular}
\end{table*}

Table~\ref{tab:ranking_metrics} demonstrates that our LLM-enhanced recommendations excel at ranking tasks, which is the main focus of our framework, as users highly value 
whether they will enjoy the recommendations or if it is even relevant to them. The LLM-enhanced versions of both SVD and SVD++ show substantial improvements across all ranking metrics. For instance, SVD-LLM achieves a $74\%$ improvement in NDCG@10 over standard SVD, and SVD++ LLM shows a $77\%$ improvement over SVD++. This confirms that our approach effectively re-ranks items according to user preferences, prioritizing items more likely to interest users even if their exact ratings might differ from the predicted rating. Furthermore, the LLM-enhanced SVD++'s performance also surpass the performance metrics of FastAI's recommendation system~\cite{FastaiMakingNeural} (e.g., our MAP being $3.69$x higher), which demonstrates the high quality of the recommendations.

\section{Related Work}

Recent advancements in Large Language Models (LLMs) have created new opportunities for enhancing recommendation systems. This section reviews relevant literature on LLM integration with traditional recommendation algorithms, focusing on approaches that align with our research direction.

\smallskip
\noindent
\textbf{LLMs for Recommendation Systems. }
Traditional recommendation algorithms like SVD have been the foundation of recommendation systems for years~\cite{koren2009matrix}. While these collaborative filtering approaches effectively capture user-item interactions, they struggle with the ``cold start'' problem and cannot easily incorporate textual user preferences. Our work bridges this gap by leveraging LLMs to enhance these traditional algorithms rather than replacing them.

Recent work by Zhang et al.~\cite{zhang2024llmtreerec} offers an innovative approach to address recommendation challenges using LLMs following the LLM as RS paradigm. Their LLMTreeRec framework tackles the system cold-start problem by structuring items into a hierarchical tree to make LLM-based recommendations more computationally efficient. 


\smallskip
\noindent
\textbf{In-context Learning for Recommendations.} Our approach employs in-context learning, where LLMs learn from demonstrations within the prompt without parameter updating~\cite{brown2020language}. According to Wu et al. \cite{wu2024survey}, prompting approaches for LLMs in recommendation systems have been widely studied, however relatively few studies have explored in-context learning for recommendation tasks. Hou et al.~\cite{hou2024large}, while they also explored zero-shot capabilities, demonstrated how in-context learning can enhance recommendation performance by providing demonstration examples. They augmented input interaction sequences by pairing prefixes with corresponding successors to create examples that improved the model's understanding of the recommendation task. Sanner et al.~\cite{sanner2023large} evaluated different in-context learning strategies for recommendation tasks, using a novel dataset that combines natural language and item-based preferences. They found that few-shot prompting outperformed other strategies, with LLMs providing competitive recommendation performance in near cold-start scenarios, relying solely on natural language preferences. Their findings inspired a shift in our approach, transitioning from natural language preferences alone to a combined item and language-based method to improve movie similarity scoring, based on user preferences for certain genres and characteristics. Liu et al.~\cite{liu2023chatgpt} evaluated ChatGPT's performance on five recommendation tasks, finding that few-shot prompting significantly outperformed zero-shot methods and traditional algorithms like Matrix Factorization and MLP, particularly in rating prediction. Their results highlighted the effectiveness of in-context learning in transferring knowledge to recommendation tasks without task-specific training. Building on these insights, our approach uses structured prompts with few-shot examples to improve similarity scoring tasks in recommendation judgments.

\smallskip
\noindent
\textbf{Enhancing Traditional Algorithms with
LLMs. }Several researchers have explored combining traditional recommendation algorithms with LLMs. Wei et al.~\cite{wei2024llmrec} proposed LLMRec, a graph augmentation framework that enhances recommender systems through three strategies revolving around enhancing the data fed to the underlying recommender algorithm: reinforcing user-item interaction edges using LLM-based sampling, enhancing item attributes, and conducting user profiling based on historical interactions. Their approach maintains data quality through specialized denoising mechanisms, significantly improving collaborative filtering performance. Similarly, Ren et al.~\cite{ren2024representation} proposed RLMRec, a model-agnostic framework that enhances existing recommenders by aligning LLM-generated semantic representations with collaborative filtering embeddings. While these approaches share conceptual similarities with our work, we focus specifically on re-ranking traditional algorithm outputs based on text preferences rather than augmenting the underlying collaborative filtering recommendation structure. Researchers may also consider initially using an LLM as RS approach which directly creates recommendations like \cite{sanner2023large} or \cite{zhang2024llmtreerec} to address the cold start problem and then transition to a collaborative filtering hybrid approach as more data is collected.

\smallskip
\noindent
\textbf{LLM-based Description Generation and Similarity Scoring. }
Acharya et al.~\cite{AcharyaLLM2023} demonstrated the effectiveness of LLM-generated item descriptions as feature inputs for recommendation systems. They used Alpaca-LoRA to generate movie and book descriptions based solely on titles, converted these to embeddings using BERT, and fed them into a GRU-based sequential recommendation model alongside item ID embeddings. 
Our framework builds on this concept by using LLMs to generate movie descriptions when official ones are unavailable, similarly leveraging these descriptions to enhance recommendation quality. 

Wang and Lim~\cite{wang2023zero} explored using LLMs for zero-shot next-item recommendation by designing prompts to generate user preference descriptions from rated items. While they ultimately use these preferences differently (their approach uses preferences to guide GPT-3 in making direct recommendations, ours uses preferences for similarity scoring to re-rank traditional algorithm recommendations), the core technique of preference extraction is similar. 
Our work builds upon this concept by automatically generating preference summaries for users in the MovieLens dataset based on their rated movies, which are then leveraged for similarity scoring and recommendation refinement. 
Maragheh et al.~\cite{yousefi2024llm} utilized LLMs to generate aspect embeddings for refining recommendations by providing contextual information and aligning with user intentions. This approach inspired our similarity scoring system, though we focus on direct similarity computation between user preferences and movie descriptions rather than aspect embedding generation.

\smallskip
\noindent
\textbf{Addressing Biases and Evaluation Challenges. }
The literature identifies several challenges in LLM-enhanced recommendation systems. These include position bias, where LLMs prioritize items in the top order~\cite{wu2024survey}, popularity bias, and fairness issues related to sensitive attributes. We address position biasing by considering each movie independently, this prevents the LLM from providing more or less attention to any specific movie. Our system also attempts to mitigate popularity bias by incorporating normalized popularity scores, popularity preferences, rating preferences, and release date preferences into the similarity calculation. The way in which we specifically incorporate popularity preferences in prompting is conceptually similar to the ``popularity bias mitigation and minimization via prompting" strategies used by \cite{lichtenberg2024largelanguagemodelsrecommender}, although our technical implementations may differ. Finally to address the fairness bias, we provide all information necessary in a single prompt for the LLM to make an informed decision, reducing its dependency on any pretrained knowledge that it may possess for any specific movie.

In addition, researchers have noted challenges in controlling LLM outputs and defining appropriate evaluation metrics. Wang et al.~\cite{wang2023rethinking} demonstrated that suitable demonstrations can effectively control the output format and content of LLMs, which directly improves recommendation evaluation metrics. Following this insight, we address these challenges by using carefully designed few-shot prompts, a retry mechanism, and regex expression searching to extract similarity scores and employing both hit rate and cumulative hit rate metrics alongside perceived quality assessments.

Finally, our work falls under the category of Generative LLMs in Non-tuning - In-context Learning as defined by~\cite{wu2024survey}, as we provide demonstration examples in our prompting, following the few-shot learning strategy. This approach allows us to leverage pre-trained LLMs without fine-tuning while still achieving high quality personalized recommendations that incorporate users' textual preferences.

\textbf{Ablation Study. }
We conducted a comprehensive ablation study to understand the contribution of each framework component by dropping them individually and measuring the impact (avg. metric across LLM-enhanced SVD and SVD++). Our ablation study confirms the critical role of natural language components in our model. Removing user input and movie descriptions resulted in the largest performance drop - 10.78\% in NDCG@10 and 13.87\% in MAP@10 - validating our central hypothesis that natural language understanding substantially enhances recommendation quality. Temporal preferences (user’s preferred release date range) were the next most influential, with 5.10\% (NDCG@10) and 6.92\% (MAP@10) declines, highlighting their importance in personalization. Popularity and rating-based features (scores and preferences) had a moderate impact (2.19\% in NDCG@10 and 3.55\% in MAP@10), contributing to recommendation diversity and quality signals, respectively. Finally, explicitly listed favorite movies showed minimal additional effect (<1\% decrease), likely due to their implicit representation in generated user preferences.

\section{Conclusion and Future Work}
In this paper, we have proposed a novel LLM-enhanced recommendation system that uses SVD and SVD++ recommendations as a foundation and fine-tunes them using a scoring mechanism provided to an LLM. Our hit rate metrics demonstrate consistent improvements across all tested $N$ values. Additionally, both SVD LLM and SVD++ LLM surpass their base algorithms in all ranking metrics proving that such LLM-enhanced recommendations improve the baseline recommendations in general. 

Our current implementation already incorporates several key features we identified as important through our testing:
\begin{enumerate}
  \item Automatic generation of user preference profiles based on favorite movies
  \item Integration of normalized popularity scores to recommend more recognizable titles
  \item Consideration of rating preferences to highlight quality content
  \item Release date preferences to match users' preferred time periods
  \item Regex expression searching to extract similarity scores accurately
  \item Few-shot prompting to teach the model the task and ensure consistent outputs even for smaller models 
\end{enumerate}

We use Phi-4-Q6\_K quantization for sophisticated language understanding while still running efficiently on a RTX 4090 Laptop GPU with 16 GB VRAM. The LLM fuzzy matching and description generation capability of our system for handling outdated or broken IMDb IDs has significantly improved the robustness of our system.

Future work will focus on further refining the automatic preference generation algorithm and optimizing the system for application on streaming platforms. We would like to study the influence of the number of demonstration examples on recommendation performance. For similarity scoring, we could implement parallelization across multiple GPUs to evaluate multiple user-movie pairs simultaneously, and batch processing to compute multiple movies' similarity scores for a single user in one LLM API call, significantly reducing API overhead. 
Similarly, for user preference profile generation, we could parallelize the process across multiple workers to generate multiple users' preference profiles simultaneously. 

Additionally, researchers may wish to experiment with collecting more user data like favorite directors, actors and genres, along with more movie data like directors, cast, genre, reviews, etc. Including this additional data in movie description profiles during similarity score generation may lead to even more nuanced understanding and improved metric results. Overall, the presented framework is designed to be applied on an as-needed basis, allowing streaming service platforms to personalize recommendations for users who wish to provide their preferences while understanding individual users better through the LLM framework.


\bibliographystyle{acm}
\bibliography{references}

\end{document}